\documentclass[conference,10pt]{IEEEtran}
\usepackage{graphics}
\usepackage{graphicx}
\usepackage[compress]{cite}
\usepackage[cmex10]{amsmath}
\usepackage{amssymb}
\usepackage{subfigure}
\usepackage{balance}
\usepackage[colorlinks=true,linkcolor=blue,citecolor=blue,urlcolor=blue]{hyperref}
\begin{document}
\title{Probing Tissue Multifractality Using Wavelet based Multifractal Detrended Fluctuation Analysis: Applications in Precancer Detection}
\author{\IEEEauthorblockN{Jalpa Soni\IEEEauthorrefmark{1},
Gregor P. Jose\IEEEauthorrefmark{2},
Sayantan Ghosh\IEEEauthorrefmark{3}, 
Asima Pradhan\IEEEauthorrefmark{4},
Tapas K. Sengupta\IEEEauthorrefmark{2},
Prasanta K. Panigrahi\IEEEauthorrefmark{1} and \\
Nirmalya Ghosh\IEEEauthorrefmark{1}\IEEEauthorrefmark{5}}

\IEEEauthorblockA{\IEEEauthorrefmark{1}Dept. of Physical Sciences, Indian Institute of Science Education and Research Kolkata (IISER-K),\\
B C K V Main Campus, P.O. Mohanpur 741 252, India.}
\IEEEauthorblockA{\IEEEauthorrefmark{2}Dept. of Biological Sciences, Indian Institute of Science Education and Research Kolkata (IISER-K),\\
B C K V Main Campus, P.O. Mohanpur 741 252, India.}
\IEEEauthorblockA{\IEEEauthorrefmark{3}School of Physics, University of KwaZulu-Natal, Private Bag X54001, Durban 4000, South Africa.}
\IEEEauthorblockA{\IEEEauthorrefmark{4}Indian Institute of Kanpur (IIT-K), Kanpur 208 017, India.}
\IEEEauthorblockA{\IEEEauthorrefmark{5}Corresponding author, EMAIL: nghosh@iiserkol.ac.in}}
\maketitle
\begin{abstract}
The refractive index fluctuations in the connective tissue layer (stroma) of human cervical tissues having different grades of precancers (dysplasia) was quantified using a wavelet-based multifractal detrended fluctuation analysis model. The results show clear signature of multi-scale self-similarity in the index fluctuations of the tissues. Importantly, the refractive index fluctuations were found to be more anti-correlated at higher grades of precancers. Moreover, the strength of multifractality was also observed to be considerably weaker in higher grades of precancers. These results were further complemented by Fourier domain analysis of the spectral fluctuations.
\end{abstract}
\maketitle

\section{Introduction}
Over the last fifty years, there have been tremendous advances in our understanding of the molecular and cellular processes of cancer; however, it still remains to be the deadliest disease of our time. Despite the significant progress made in treatment of a number of neoplastic disorders, early detection of neoplastic changes appears to be our best method to improve patient quality of life and reduce cancer mortality. The conventional methods for early detection and diagnosis of cancer rely on histological and cytological examination of tissue. This approach rely on characterizing morphologic and architectural alterations, including increased nuclear size, increased nucleus/cytoplasm ratio, hyperchromasia, pleomorphism, and loss of normal epithelial architecture. There is a tremendous need in developing better methodologies for extraction and quantification of the morphological alterations associated with cancer/pre-cancer development. In this regard, the optical spectroscopic and imaging approaches have shown early promise in quantifying both the morphological (using elastic scattering spectroscopy) and biochemical (using in-elastic scattering spectroscopy such as fluorescence and Raman) alterations associated with cancer development \cite{ramanujam,boustany,alfano}. Notably, polarized elastic scattering spectroscopy has been explored to quantify the self-similar (fractal) nature of micro-scale fluctuation of local refractive index in tissues. These studies have revealed that changes in tissue self-affinity can serve as a potential bio-marker for pre-cancer \cite{hunter}.
\par
Since most of the cancers arise in epithelial tissues, majority of the previous attempts on developing methods (optical or non-optical) for early diagnosis of cancer relied largely on quantifying the alterations in the superficial epithelial layer. However, recently it has been recognized that in addition to the alterations in the superficial epithelial cells, neoplasia is also associated with characteristic changes in the underlying connective tissue layer (stroma). Progression of cancer involves complex interactions between neoplastic cells and the stroma. Also, carcinogenesis results, in part, from defective epithelial-stromal communication \cite{pupa123}.  In fact, alterations in stromal biology may precede and stimulate neoplastic progression in pre-invasive disease \cite{herlyn}. Interestingly, the collagen fiber network present in stroma also exhibits fractal architecture in the organization of the fibers and micro-fibrils \cite{arifler}. Quantification of the changes in the fractal characteristics of the stroma may thus provide additional targets to aid in screening and early detection of precancerous changes. With this motivation, we have explored the use of a wavelet-based multifractal detrended fluctuation analysis model to extract and quantify the fractal properties of the refractive-index structure (inhomogeneities) in the stromal layer of dysplastic (pre-cancerous) human cervical tissues. The results showed interesting variations in the micro-optical tissue fractal properties in the various states of dysplasia. The details of these results are presented in this paper. 

\section{EXPERIMENTAL METHODS AND MATERIALS}
We have used a phase contrast microscope to measure the spatial fluctuations of the refractive index structures of the tissues. The samples used in this study were pathologically characterized, unstained sections of human cervical tissues (containing both epithelial and stromal regions) in glass slide. Tissue specimens were frozen and vertically sectioned (epithelium to stromal region) to a thickness of $5\mu m$ for measurements and analysis. The lateral dimensions of tissue sample were typically $4 mm\times6 mm$. The dysplastic tissues were histopathologically characterized as Grade I, II and III respectively, based on their epithelial architecture.
\par
The phase contrast images were recorded from the stromal region of the tissue sections using a phase contrast microscope (Olympus IX81, USA).  Images of the specimens were taken at a magnification of 100X and were recorded with a CCD camera (ORCA-ERG, Hamamatsu) having $1344\times1024$ pixels (pixel dimension $6.45 \mu m$).
\section{Theory}
Biological tissue is an optically inhomogeneous medium having random fluctuations in local refractive indices (arising from the presence of microscopic inhomogeneities; macromolecules, cell organelles, organized cell structure, extra-cellular matrix, interstitial layers etc.), the spatial scale of which varies from few dozen nanometers to several tens of micrometers. Tissue can thus be described as a random continuum of the inhomogeneities of the refractive index with varying spatial scale. As noted previously, for many types of tissues, the spatial distribution of refractive index have been found to exhibit statistical self-similarity \cite{hunter,schmitt}. Specifically, the stromal tissues (investigated in this study) are known to be comprised of complex fibrous network having fractal-like arrays of collagen molecule, micro-fibrils and fiber bundles \cite{arifler}. Since, the phase contrast image is a direct indicator of such spatial fluctuation of local refractive index structure, analysis of this fluctuation through appropriate statistical models may yield wealth of interesting micro-optical parameters, each of which can potentially serve as a useful biological metric. In order to explore this possibility, we have first unfolded (in one direction) the recorded images to obtain statistically large enough refractive index fluctuation series for analysis. These were then analyzed through (i) Fourier analysis and (ii) Wavelet based Multi Fractal De-trended Fluctuation Analysis (WB-MFDFA). 
\subsection{Fourier Domain Analysis}
Fourier domain analysis is a simple and convenient approach to test and quantify the fractal behavior of any fluctuation series. Such analysis based on the power spectrum of the Fourier transform of fluctuations has been widely used in many areas dealing with noise, from chaos to order transitions. Recently this approach has also been used to analyze light scattering from biological tissues \cite{hunter,schmitt}. The power spectrum can be related to the Hurst scaling exponent (H, a measure of the fractal nature: $ 0 \leq H \leq1$ for fractals, ) of any (one dimensional) fluctuation series exhibiting statistical self-similarity by the following relationship \cite{hurst}
\begin{eqnarray}
P(k)\approx k^{-\alpha} \nonumber \\
\alpha=2H+1
\label{eq:eq1}
\end{eqnarray}
Here, in the power spectrum $P(k)$, $k$ is the frequency (spatial frequency in our case).
\par
Note that Eq. \ref{eq:eq1} is based on the assumption of an exact power law autocorrelation function of the fluctuations. In fact, such dependence can be understood by noting that the power spectrum of any fluctuation series is related to the autocorrelation function through an inverse Fourier transform. The exact form of the power spectrum is different though for fluctuations having more general type of autocorrelation functions (such as the von Karman self-affine function, which has been extensively used to describe refractive index correlations in tissues and other similar type of random media \cite{hunter,schmitt}). Never-the-less at the limit of large spatial frequencies, the power spectrum for those also essentially converge to Eq. \ref{eq:eq1} \cite{schmitt}. 
\par
In the first step, we have thus analyzed the recorded tissue refractive index fluctuations (one dimensional) employing discrete Fourier transform (DFT) and its power spectrum. The obtained power spectrum was fitted to Eq. \ref{eq:eq1} to yield the value for the exponent $\alpha$ and the Hurst scaling exponent $H$. 
\par
Note that the above analysis is based on monofractal hypothesis, which assumes that the scaling properties are the same for the entire region of the tissue section examined. However, considering the wide range of the dimensions of the inhomogeneities (in refractive index) and the complex nature of the correlations present between them in tissue (the stromal tissues), this assumption may be unrealistic. In such situation, it would be desirable to use a more general type of analysis which could extract and quantify the nature of multi-fractality (if there) present in the signal. A multifractal signal can be decomposed into many subsets, wherein a local Hurst exponent quantifies the local singular behavior and gives the local scaling of the signals. Such multifractal analysis may provide additional diagnostic information on the nature of the nonlinearity encoded in the Fourier space of the refractive index fluctuations. We therefore applied a more general wavelet based approach on the multifractal detrended fluctuation analysis (WB-MFDFA), as discussed below.
\subsection{Wavlet Based Multi Fractal Detrended Fluctuation Analysis (WB-MFDFA)}
The multi fractal detrended fluctuation analysis (MFDFA) has been introduced to study the long range correlations in fluctuations when a trend is present \cite{peng}. It has been found to be very useful for studying non-stationary fluctuation series, where it is of vital importance to separate local fluctuations from average behavior (trend). In MFDFA and its variants, the detrending is usually done using polynomial fits. However, a more recent approach exploits the use of discrete wavelets for performing the detrending procedure \cite{mani,mani4}. It has been shown that the natural, built-in variable window size in wavelet transforms makes this procedure more efficient for this purpose. We have therefore explored the use of this Wavelet based Multi Fractal Detrended Fluctuation Analysis (WB-MFDFA) for analyzing our phase contrast data from tissue sections \cite{mani}. 
\par
In this approach, the profile is first calculated from a series $x_t(t=1,\ldots,N)$ of length N, by performing the cumulative sum of the series after subtracting the mean.
\begin{equation}
Y(i)=\sum_{t=1}^{i}[x_t-\langle x \rangle],\quad i=1,\ldots,N.
\end{equation}
In order to separate the fluctuation from the trend, a discrete wavelet transform is then carried out on the profile $Y(i)$. Discrete wavelets belonging to Daubechies family (Db4) is used for this purpose. It may be worth mentioning here that since, these wavelets satisfy the vanishing moment conditions, the low-pass coefficients keep track of the polynomial trend in the data. Thus, reconstruction using the low-pass coefficients alone is quite accurate in extracting the local trend, in a desired window size. The fluctuations are then extracted at each level by subtracting the obtained series from the original data. The details of this detrending procedure can be found in \cite{mani}. Thus obtained fluctuations are then considered for further analysis, as follows. The extracted fluctuations are subdivided into non-overlapping segments $M_s=\textrm{int}(N/s)$ where the window size at a particular level $(L)$ is, $s=2^{(L-1)}W$ for the chosen wavelet. $W$ is the number of filter coefficients of the discrete wavelet transform under consideration. This procedure is repeated from the end to beginning to calculate the local variance in case $N/s$ is not an integer which would mean that some data points have to be discarded. Following this procedure, the $q^th$ order fluctuation function, $F_q(s)$ is obtained as,
\begin{equation}
F_q(s)\equiv\left\{\frac{1}{2M_s}\sum_{b=1}^{2M_s}\left[F^2(b,s)\right]^{q/2}\right\}^{1/q}
\label{eq:eq3}
\end{equation}					
Here `$q$' is the order of the moment which can take any real value. This procedure is repeated for various values of $q~(q\neq0)$ for variable window sizes (in our case we varied $-10 \leq q \leq 10$). The scaling behavior of the fluctuations can be obtained by analyzing the fluctuation function in a logarithmic scale for each value of $q$
\begin{equation}
F_q(s)\sim s^{h(q)}
\end{equation}
Note that special care must be taken for a value of $q=0$. This follows because for $q=0$, direct evaluation of Eq. \ref{eq:eq3} leads to divergence of the scaling exponent \cite{mani}.  In that case, logarithmic averaging has to be employed to find the fluctuation function 
\begin{equation}
F_q(s)\equiv\exp \left\{\frac{1}{2M_s}\sum_{b=1}^{2M_s}\log\left[F^2(b,s)\right]^{q/2}\right\}^{1/q}
\end{equation}
The generalized Hurst exponent $h(q)$ values are independent of $q$ if the series is mono-fractal. For multifractal series, $h(q)$ values depend on $q$. Note that the value for $h$ at $q = 2$, corresponds to the Hurst exponent $(H = h(q = 2))$ for a stationary series. In fact, the correlation behavior of any series (stationary or non-stationary) is characterized from the Hurst exponent. For long range correlation, $H > 0.5$, $H = 0.5$ for uncorrelated and $H <0.5$ for long range anti-correlated \cite{hurst}. The variation $h(q)$ can thus be used as an indicator to correlation behavior and multifractality of the fluctuations.  
\section{Results and discussions}
Typical phase contrast images recorded from the stromal region of dysplastic cervical tissues, histopathologically characterized   Grade I and Grade III, are shown in Figure 1a and 1b respectively. The wide range of sizes and shapes of the index inhomogeneities and their high packing densities are evident. The complex nature of the index variation however, underscores the problem– how does one extract and quantify useful diagnostic metrics from this intertwined information? The observed index variations were then unfolded to generate the fluctuation series, which were subsequently analyzed via the Fourier analysis and WB-MFDFA.
\begin{figure}[h]
\begin{center}
\subfigure[]{\label{fig:imagegr1}
\includegraphics[width=6.5cm,height=6.5cm]{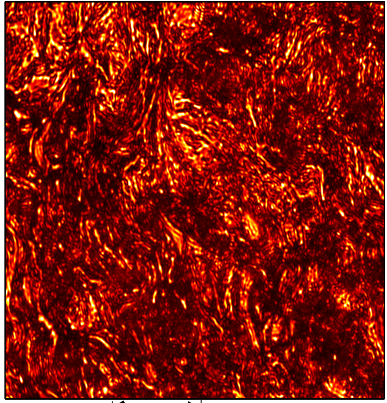}
}
\subfigure[]{\label{fig:imagegr3}
\includegraphics[width=6.5cm,height=6.5cm]{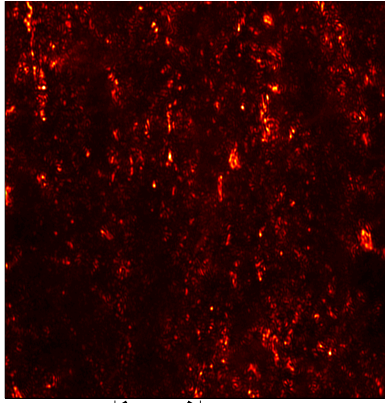}
}
\caption{\label{fig:images}\scriptsize{(Color Online) Phase-contrast images of the stromal region of cervical tissues having \subref{fig:imagegr1} Grade I and \subref{fig:imagegr3} Grade III stage of dysplasia.}}

\end{center}
\end{figure}
\subsection{Results of the Fourier analysis}
\begin{figure}[h]
\begin{center}
\subfigure[]{\label{fig:fftgr1}
\includegraphics[width=6.5cm,height=6.5cm]{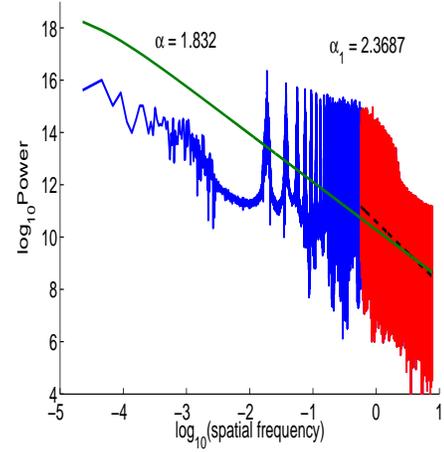}
}
\subfigure[]{\label{fig:fftgr3}
\includegraphics[width=6.5cm,height=6.5cm]{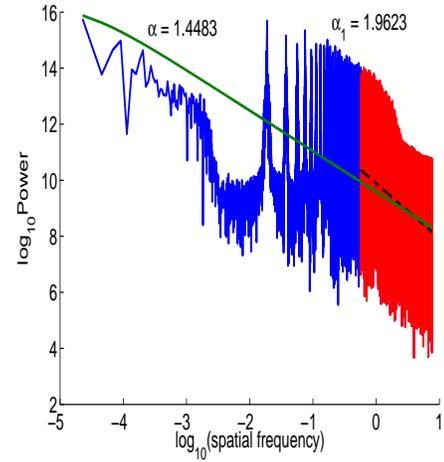}
}
\caption{\label{fig:fftgr}\scriptsize{(Color Online) Power spectra of index variations in the stromal region of cervical tissues having \subref{fig:fftgr1} Grade I and \subref{fig:fftgr3} Grade III stage of dysplasia. The power spectral density and the spatial frequency are displayed in a log-log scale.}}

\end{center}
\end{figure}

Figure \ref{fig:fftgr1} and \ref{fig:fftgr3} show the corresponding power spectral density of the index inhomogeneities in the two tissues, which we obtained by performing Fourier transformation on the unfolded fluctuation series. The power spectral density and the spatial frequency are displayed in a log-log scale in these figures. 
Several interesting characteristics of the spectrum can be discerned. Beyond a certain spatial frequency range, the spectral density appears linear when plotted on a log–log scale, indicating a power-law scaling which depends on the sizes of the inhomogeneities that contribute to index variations. However, this is also associated with large background fluctuations indicating the overall randomness of the underlying index variations. Interestingly, the power law coefficient does not appear to be uniform throughout the entire spatial frequency region. This behavior is a clear indication of the multifractal nature of the index fluctuations. The data were therefore fitted to Eq. \ref{eq:eq1} for two different selected spatial frequency range; (i) for the overall broad range of the spatial frequencies where the linear behavior was noted (power law exponent noted as $\alpha$ in the figures) and (ii) at a narrower higher spatial frequency range where the slope was apparently different (power law exponent noted as $\alpha_1$ in the figures). The results of these fits are displayed as solid and dashed lines respectively, in the figures.  The corresponding estimates for the exponents are also listed in the figures. Importantly, the value for the estimates for the average Hurst parameter are significantly different, $H = 0.416$ and $0.224$ for Grade I and Grade III dysplastic connective tissues respectively. A lower value of $H$ is indicative of increasing roughness of the medium, i,e predominance of index inhomogeneities having smaller spatial dimensions. Note that the limiting values of $H$, unity and zero correspond to a smooth Euclidean random field (marginal fractal) and a space-filling field (extreme fractal) respectively \cite{hurst}. The lower value of $H$ in higher grades of dysplasia is consistent with previous reports where it has been observed that the volume fraction of collagen fibers in the stroma decreases with precancer and cancer progression, and fibers tend to be shorter and more disconnected in neoplastic stroma \cite{arifler}.
\par
Never-the-less the observed nonlinearity in the Fourier space of the refractive index fluctuations (different values of the exponent $\alpha$ at different spatial frequency range) underscores the need for employing a multi-resolution fluctuation analysis model. 
\subsection{Results of the analysis of WB-MFDFA}
\begin{figure}[h]
\begin{center}
\subfigure[]{\label{fig:gr1hq}
\includegraphics[width=6.5cm,height=6.5cm]{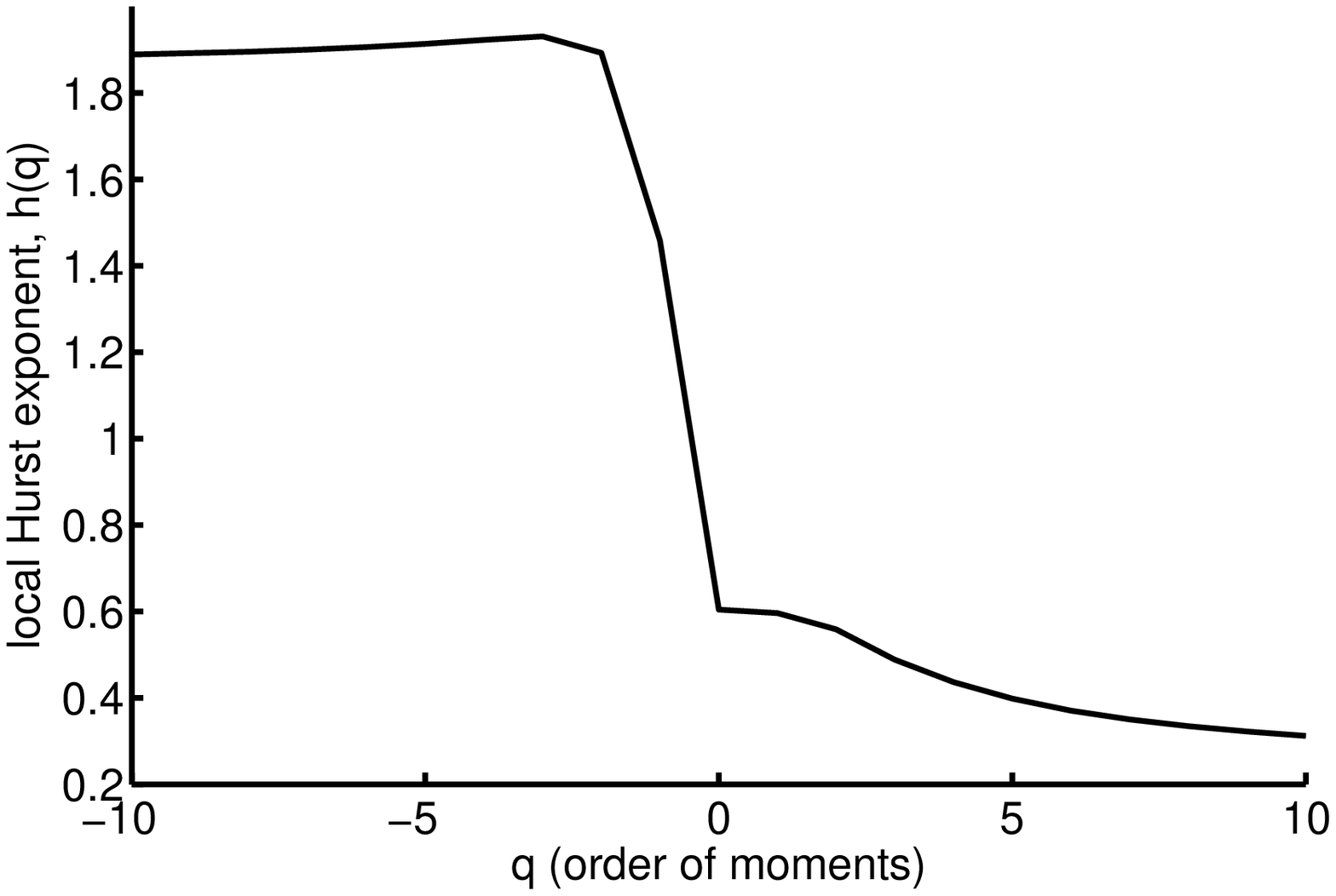}
}
\subfigure[]{\label{fig:gr3hq}
\includegraphics[width=6.5cm,height=6.5cm]{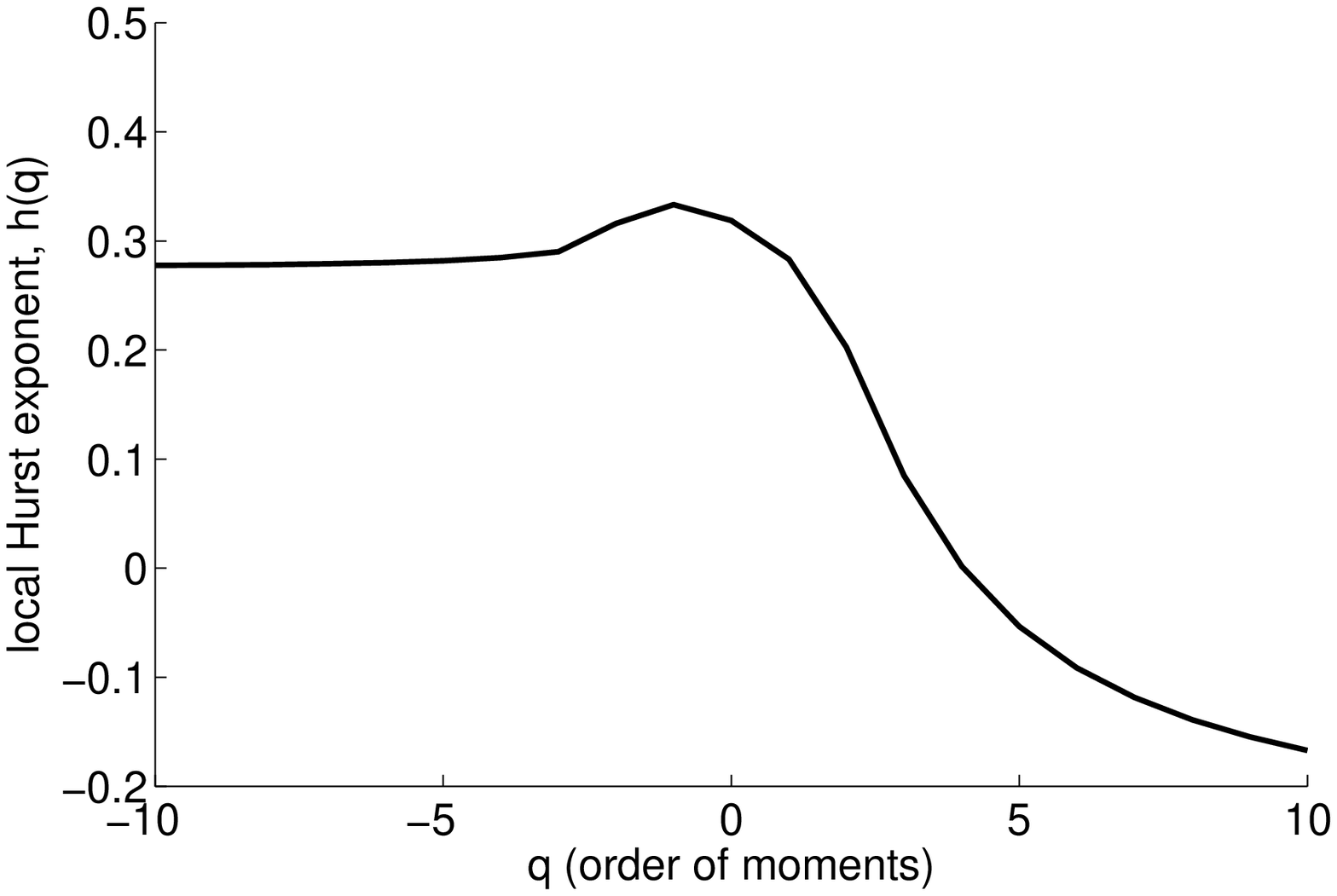}
}
\caption{\label{fig:grhq}\scriptsize{(Color Online) The variation of generalized Hurst exponent $h(q)$ as a function of order of the moment $(q)$ for the index variations in the stromal region of cervical tissues having \subref{fig:gr1hq} Grade I and \subref{fig:gr3hq} Grade III stage of dysplasia. The value for $h(q=2)$ for these two tissues were $0.55$ and $0.203$ respectively.}}

\end{center}
\end{figure}

In Figure \ref{fig:gr1hq} and \ref{fig:gr3hq}, we summarize the results of the Wavelet based Multi Fractal Detrended Fluctuation Analysis performed on the unfolded index fluctuations series in the two tissues (whose results have been presented in Figure \ref{fig:images} and \ref{fig:fftgr}). The dependence of the generalized Hurst exponent $h (q)$ on the order of the moment $(q)$ are displayed in the figures.
\par
As is apparent, the index fluctuations for both these tissues exhibit long range correlations and multifractal behavior (manifested as a variation of $h (q)$). Interestingly, the variation of $h (q)$ is observed to be considerably weaker for the Grade III dysplastic connective tissue, as compared to that for the Grade I tissue. This clearly indicates that with increasing grades of dysplasia (with precancer progression), the strength of multifractality also decreases. This was further quantified and confirmed by determining the singularity spectrum of multifractality (not shown here) \cite{peng,mani}. As expected, the width of the singularity spectrum was found to be considerably narrower in case of the Grade III tissues compared to the Grade I tissues, indicating a trend towards monofractal behavior. The values for $h (q = 2)$ for these two tissues (Grade I and III) were determined to be $0.55$ and $0.203$ respectively. Note that these values should correspond to the Hurst exponent $(H = h(q = 2))$ for a stationary series. The observed variations in the values for these parameters (in the two tissues) are qualitatively similar to that observed for the Hurst exponent $(H)$ determined using the Fourier analysis (in both cases, the value for the Hurst exponent decreases at higher grade of precancer indicating a trend towards long range anti-correlation). The absolute values estimated using these two approaches are different. This is expected because, the Fourier analysis estimates this parameter $(H)$ using a monofractal approximation that is far from being true for these tissues (as is apparent from the non-stationary nature of the fluctuations). Importantly, none of the multifractal trends observed in these tissues could be gleaned using the Fourier analysis, where at best one can qualitatively predict about the overall nature of the self-similar behavior of the tissue refractive index fluctuations. Derivation and quantification of the multifractal trends is possible due to the multi-resolution ability of the WB-MFDFA approach. 

\section{CONCLUSIONS}
To conclude, we have employed a wavelet based approach on the multifractal detrended fluctuation analysis (WB-MFDFA) to extract and quantify self-similarity of tissue refractive index fluctuations. This novel approach for fluctuation analysis has been explored here in a scenario of significant clinical interest, that for detecting precancerous alterations in human cervical tissues. Early indications show promise of this approach for extraction and quantification of the morphological alterations associated with precancers. The information obtained on the multifractal nature of refractive index fluctuations in tissue may also prove to be valuable for developing light scattering based spectroscopic and imaging techniques for non-invasive and early diagnosis of cancer. This is currently under investigation in our laboratory.  

\begin{thebibliography}{10}
\providecommand{\url}[1]{#1}
\csname url@samestyle\endcsname
\providecommand{\newblock}{\relax}
\providecommand{\bibinfo}[2]{#2}
\providecommand{\BIBentrySTDinterwordspacing}{\spaceskip=0pt\relax}
\providecommand{\BIBentryALTinterwordstretchfactor}{4}
\providecommand{\BIBentryALTinterwordspacing}{\spaceskip=\fontdimen2\font plus
\BIBentryALTinterwordstretchfactor\fontdimen3\font minus
  \fontdimen4\font\relax}
\providecommand{\BIBforeignlanguage}[2]{{%
\expandafter\ifx\csname l@#1\endcsname\relax
\typeout{** WARNING: IEEEtran.bst: No hyphenation pattern has been}%
\typeout{** loaded for the language `#1'. Using the pattern for}%
\typeout{** the default language instead.}%
\else
\language=\csname l@#1\endcsname
\fi
#2}}
\providecommand{\BIBdecl}{\relax}
\BIBdecl

\bibitem{ramanujam}
\BIBentryALTinterwordspacing
N.~Ramanujam, ``Fluorescence spectroscopy of neoplastic and non-neoplastic
  tissues,'' \emph{Neoplasia}, vol.~2, no. 1--2, p.~89, 2000. [Online].
  Available: \url{http://www.ncbi.nlm.nih.gov/pmc/articles/PMC1531869/}
\BIBentrySTDinterwordspacing

\bibitem{boustany}
\BIBentryALTinterwordspacing
N.~N. Boustany, S.~A. Boppart, and V.~Backman, ``{Microscopic Imaging and
  Spectroscopy with Scattered Light},'' \emph{Annual Review of Biomedical
  Engineering}, vol.~12, no.~1, pp. 285--314, 2010. [Online]. Available:
  \url{http://dx.doi.org/10.1146/annurev-bioeng-061008-124811}
\BIBentrySTDinterwordspacing

\bibitem{alfano}
\BIBentryALTinterwordspacing
S.~P. Schantz, V.~Kolli, H.~E. Savage, G.~Yu, J.~P. Shah, D.~E. Harris,
  A.~Katz, R.~R. Alfano, and A.~G. Huvos, ``In vivo native cellular
  fluorescence and histological characteristics of head and neck cancer.''
  \emph{Clinical Cancer Research}, vol.~4, no.~5, pp. 1177--1182, 1998.
  [Online]. Available:
  \url{http://clincancerres.aacrjournals.org/content/4/5/1177.abstract}
\BIBentrySTDinterwordspacing

\bibitem{hunter}
\BIBentryALTinterwordspacing
M.~Hunter, V.~Backman, G.~Popescu, M.~Kalashnikov, C.~W. Boone, A.~Wax,
  V.~Gopal, K.~Badizadegan, G.~D. Stoner, and M.~S. Feld, ``Tissue
  self-affinity and polarized light scattering in the born approximation: A new
  model for precancer detection,'' \emph{Phys. Rev. Lett.}, vol.~97, no.~13, p.
  138102, Sep 2006. [Online]. Available:
  \url{http://dx.doi.org/10.1103/PhysRevLett.97.138102}
\BIBentrySTDinterwordspacing

\bibitem{pupa123}
\BIBentryALTinterwordspacing
S.~M. Pupa, S.~M\'enard, S.~Forti, and E.~Tagliabue, ``New insights into the
  role of extracellular matrix during tumor onset and progression,'' \emph{J.
  Cell. Physiol.}, vol. 192, no.~3, pp. 259--267, 2002. [Online]. Available:
  \url{http://dx.doi.org/10.1002/jcp.10142}
\BIBentrySTDinterwordspacing

\bibitem{herlyn}
\BIBentryALTinterwordspacing
T.~Bogenrieder and M.~Herlyn, ``Axis of evil: molecular mechanisms of cancer
  metastasis,'' \emph{{Oncogene}}, vol.~22, pp. 6524--6536, 2003. [Online].
  Available: \url{http://dx.doi.org/10.1038/sj.onc.1206757}
\BIBentrySTDinterwordspacing

\bibitem{arifler}
\BIBentryALTinterwordspacing
D.~Arifler, I.~Pavlova, A.~Gillenwater, and R.~Richards-Kortum, ``Light
  scattering from collagen fiber networks: Micro-optical properties of normal
  and neoplastic stroma,'' \emph{Biophysical Journal}, vol.~92, no.~9, pp. 3260
  -- 3274, 2007. [Online]. Available:
  \url{http://dx.doi.org/10.1529/biophysj.106.089839}
\BIBentrySTDinterwordspacing

\bibitem{schmitt}
\BIBentryALTinterwordspacing
J.~M. Schmitt and G.~Kumar, ``Turbulent nature of refractive-index variations
  in biological tissue,'' \emph{Opt. Lett.}, vol.~21, no.~16, pp. 1310--1312,
  Aug 1996. [Online]. Available: \url{http://dx.doi.org/10.1364/OL.21.001310}
\BIBentrySTDinterwordspacing

\bibitem{hurst}
H.~E. Hurst, ``{Long-term storage capacity of reservoirs},'' \emph{Trans. Amer.
  Soc. Civil Eng.}, vol. 116, pp. 770--808, 1951.

\bibitem{peng}
\BIBentryALTinterwordspacing
C.-K. Peng, S.~V. Buldyrev, S.~Havlin, M.~Simons, H.~E. Stanley, and A.~L.
  Goldberger, ``Mosaic organization of dna nucleotides,'' \emph{Phys. Rev. E},
  vol.~49, no.~2, pp. 1685--1689, Feb 1994. [Online]. Available:
  \url{http://dx.doi.org/10.1103/PhysRevE.49.1685}
\BIBentrySTDinterwordspacing

\bibitem{mani}
\BIBentryALTinterwordspacing
P.~Manimaran, P.~K. Panigrahi, and J.~C. Parikh, ``Wavelet analysis and scaling
  properties of time series,'' \emph{Phys. Rev. E}, vol.~72, no.~4, p. 046120,
  Oct 2005. [Online]. Available:
  \url{http://dx.doi.org/10.1103/PhysRevE.72.046120}
\BIBentrySTDinterwordspacing

\bibitem{mani4}
\BIBentryALTinterwordspacing
{Manimaran, P. and Panigrahi, P.K. and Parikh, J.C.}, ``{Difference in nature
  of correlation between NASDAQ and BSE indices},'' \emph{Physica A:
  Statistical Mechanics and its Applications}, vol. 387, no.~23, pp.
  5810--5817, 2008. [Online]. Available:
  \url{http://dx.doi.org/10.1016/j.physa.2008.06.033}
\BIBentrySTDinterwordspacing

\end{thebibliography}

\end{document}